\documentclass[runningheads]{llncs}
\usepackage{fullpage}
\usepackage{graphicx} 
\usepackage[T1]{fontenc}
\usepackage{wrapfig}
\usepackage[]{subcaption}

\usepackage{url}
\usepackage{tikz}
\usepackage{xcolor}
\definecolor{darkred}{rgb}{0.6, 0, 0}
\definecolor{darkgreen}{rgb}{0, 0.5, 0}
\definecolor{darkblue}{rgb}{0, 0, 0.5}
\definecolor{darkmagenta}{rgb}{0.5, 0, 0.5}
\usepackage{hyperref}
\hypersetup{
    colorlinks=true,      
    linkcolor=darkred,    
    filecolor=darkgreen,  
    urlcolor=darkblue,    
    citecolor=darkmagenta 
}

\usepackage{ifthen}

\newcommand{\imagedir}{./} 
\immediate\write18{if [ -d "./images" ]; then echo "\\\"; fi > checkdir.tex}

\InputIfFileExists{checkdir.tex}{}{}

\begin{document}

\title{Ktirio Urban Building: A Computational Framework for City Energy Simulations Enhanced by CI/CD Innovations on EuroHPC Systems}
\author{}
\author{Luca Berti\inst{1} \and Vincent Chabannes \inst{1}\orcidID{0009-0005-3602-3524} \and
Javier Cladellas \inst{1}\orcidID{0009-0003-8687-7881} \and 
Abdoulaye Diallo \inst{1}\orcidID{0009-0006-8731-0547}\ \and
Maryam Maslek Elayam \inst{1}\orcidID{0000-0003-0880-5180} \and
Philippe Pinçon \inst{1}\orcidID{0009-0009-7724-3055 } \and
Christophe Prud'homme\inst{1}\orcidID{0000-0003-2287-2961}}
\institute{Cemosis, IRMA UMR 7501, University of Strasbourg, CNRS\\ 
\email{\{vincent.chabannes,christophe.prudhomme\}@cemosis.fr}}

\authorrunning{V. Chabannes et al.}

\maketitle

\begin{abstract}
The building sector in the European Union significantly impacts energy consumption and greenhouse gas emissions. The EU's Horizon 2050 initiative sets ambitious goals to reduce these impacts through enhanced building renovation rates. The CoE HiDALGO2 supports this initiative by developing high-performance computing solutions, specifically through the Urban Building pilot application, which utilizes advanced CI/CD methodologies to streamline simulation and deployment across various computational platforms, such as the EuroHPC JU supercomputers. The present work provides an overview of the Ktirio Urban Building framework (KUB), starting with an overview of the workflow and a description of some of the main ingredients of the software stack and discusses some current results performed on EuroHPC JU supercomputers using an innovative CI/CD pipeline.


\end{abstract}

\section{Introduction}
The building sector accounts for approximately 40\% of final energy consumption and 36\% of greenhouse gas emissions within the European Union~\cite{european_commision_energy_2020}. In response, the EU has established ambitious targets under the Horizon 2050 framework to double energy renovation rates over the next decade~\cite{european_commision_stakeholder_2021}, highlighting the need for innovative solutions to drive these initiatives forward. The Centre of Excellence (CoE) HiDALGO2 project, focusing on high-performance computing and advanced simulations, is at the forefront of tackling this challenge, mainly through its Urban Building pilot application.

The Ktirio Urban Building (KUB) pilot in CoE HiDALGO2 aims to leverage high-performance computing to enhance city energy simulation for better energy management and air quality assessment. Advanced simulation tools predict energy consumption, thermal comfort, and indoor air quality across both the building and urban scales. These simulations support detailed individual building-level analysis and extend to broader urban environments, influencing urban planning and policy-making. KUB is part of the platform Ktirio~\cite{cemosis_ktirio_2024} which itself is based on Feel++~\cite{christophe_prudhomme_feelppfeelpp_2024}.

Building on the foundations of the Feel++ framework, our project introduces an innovative CI/CD environment specifically designed for deployment on EuroHPC JU supercomputers. This environment enhances our ability to develop and test simulations rapidly and provides a robust platform for deploying these simulations at scale. Integrating EuroHPC JU infrastructures represents a significant advancement in making high-performance computing resources accessible for urban simulation studies.

Thanks to synergies with another CoE HiDALGO2 pilot, the KUB project is set to integrate building simulations with urban air pollution (UAP) models to comprehensively assess the environmental impact of building stocks soon. This integration improves the predictive accuracy of the simulations by incorporating real-time data such as wind speed and solar radiation, enhancing the models' responsiveness to environmental conditions.

Implementing these objectives will facilitate more informed urban planning decisions, support policy development for energy efficiency, and contribute to reducing urban greenhouse gas emissions. The project also focuses on enhancing the interaction between different environmental models to provide a holistic view of urban ecosystems.

The paper is organized as follows: Section 2 discusses KUB's current workflow, Section 3 provides an overview of some modeling and simulation components, and finally, Section 4 delves into the CI/CD processes from standard DevOps to HPCops tailored for EuroHPC environments.


\section{Ktirio Urban Building Workflow}

\begin{figure}
    \centering
    \includegraphics[width=.9\textwidth]{\imagedir 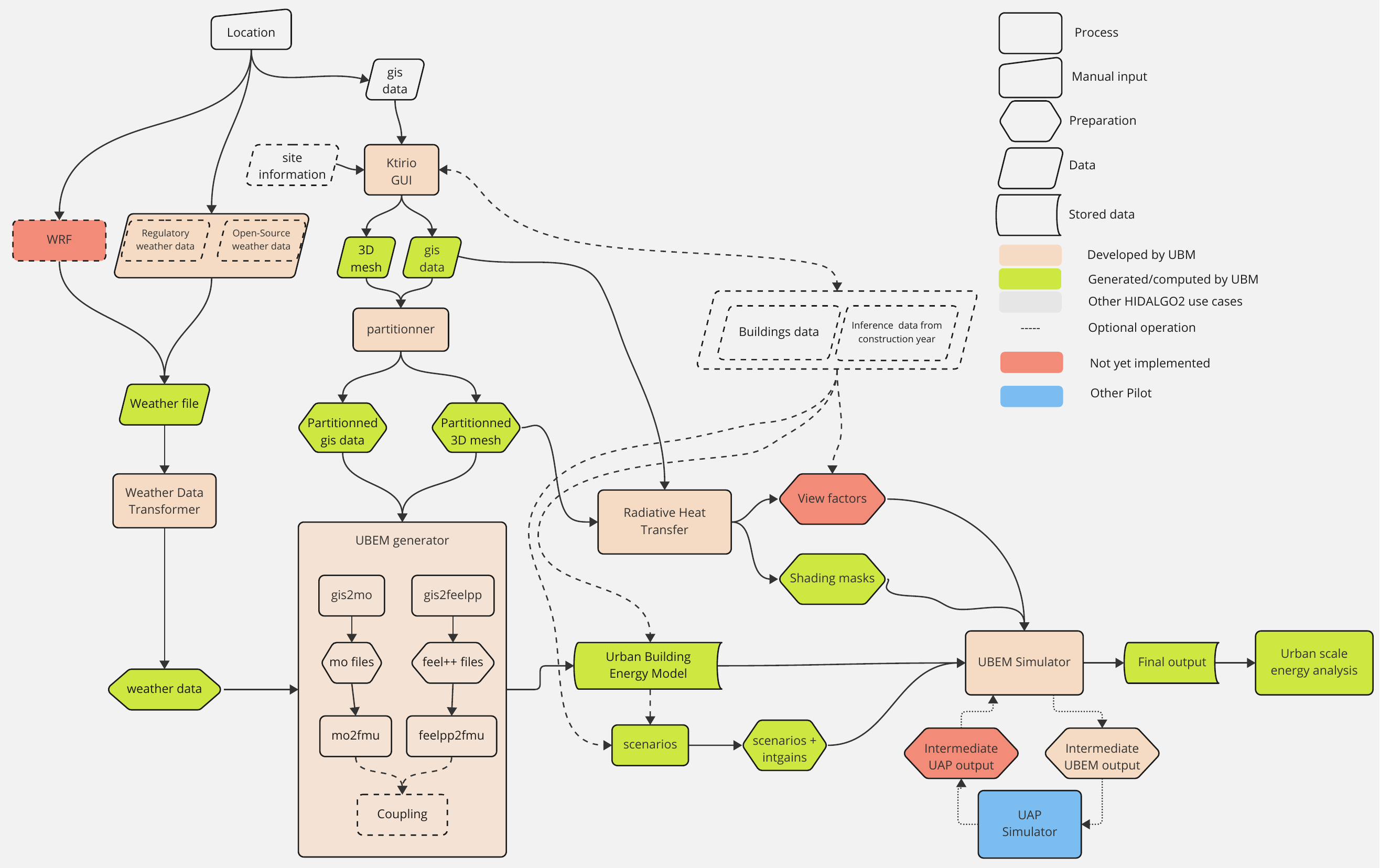}
    \caption{Current Urban Building Workflow from localization to city energy simulation report.}
    \label{fig:kub-workflow}
\end{figure}

The Ktirio urban building workflow, see Figure~\ref{fig:kub-workflow}, integrates various data sources and computational tools to simulate and analyze urban building energy and its impact on urban environments. The process encompasses data acquisition, processing, simulation, and analysis, eventually coupled with urban air pollution (UAP) models.

\subsection{Data Handling and Simulation Process}
The workflow begins with collecting and preparing GIS and weather data, transforming it into a format usable for simulations. This data is then partitioned for scalable processing and converted into Modelica and Feel++\cite{christophe_prudhomme_feelppfeelpp_2024} compatible formats through the UBEM Generator.

This processed data is employed to simulate energy consumption and indoor environmental quality using the Urban Building Energy Model (UBEM). The simulation focuses on radiative heat transfer, enhancing the accuracy of the energy models. It also computes view factors and shading masks, assessing how buildings affect each other's exposure to natural light and heat, influencing the urban heat island effect and overall building energy needs.

The building simulation outputs are then optionally fed into the urban air pollution (UAP) simulator to evaluate the impact of building emissions on urban air quality. A feedback loop refines the building simulation scenarios based on intermediate outputs from the UAP simulator, ensuring that the models accurately reflect the complex interdependencies between urban building energy usage and urban air quality.

\subsection{Final Analysis and Urban Scale Energy Evaluation}
The final step involves the UBEM Simulator, which generates large-scale outputs that summarize the overall energy consumption and environmental impact of buildings on an urban scale. This comprehensive urban scale analysis merges data from the building energy and air quality models to provide a holistic view of urban environmental quality.

This streamlined workflow is critical for accurately simulating and understanding urban sustainability challenges, supporting our application's broader objectives of improving urban living conditions and environmental impact.

\section{Overview of Urban Building Modeling and Simulation}

We now provide an overview of the geometrical and physical modeling and simulation components of the Urban building application.

\subsection{Geometry Reconstruction of the KUB Urban Model}

The geometric reconstruction of urban environments within KUB involves a sophisticated approach to creating multi-fidelity representations of buildings, terrain, vegetation, roads, and other urban elements. This section outlines the methodologies employed and the various levels of detail (LOD) used in the models.

The primary challenge lies in accurately representing the complex urban landscape to support various simulations. A tiled web map approach is adopted, which allows for distributed data management and adapts the Level of Detail (LOD) based on specific needs. However, this approach requires carefully integrating tiles to ensure seamless representation.

We describe our definition of Levels of Detail for Buildings such as:
\begin{itemize}
    \item \textbf{LOD-0:} The most straightforward form, representing buildings as oriented bounding boxes. This level is typically used for large-scale preliminary analyses and quick visual assessments.
    \item \textbf{LOD-1:} Buildings are represented as polygonal extrusions, with added roof structures to improve the visual accuracy and utility in simulations that do not require detailed internal features.
    \item \textbf{LOD-2:} At this level, buildings are detailed using Industry Foundation Classes (IFC) standards, supporting detailed thermal and structural analyses. This includes detailed geometries for each building entity, often using complex surface models like B-REP, swept solids, or CSG techniques.
\end{itemize}

\begin{wrapfigure}{R}{0.6\textwidth} 
\centering
\subfloat[LOD-0: a building is represented by its bounding box]{%
  \includegraphics[width=0.48\linewidth]{\imagedir 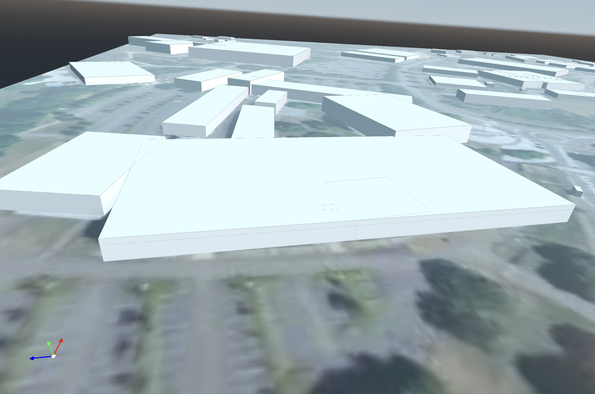}
  \label{fig:building-lod0}
}
\subfloat[LOD-1: a building is represented by its ground footprint elevated to its height]{%
  \includegraphics[width=0.48\linewidth]{\imagedir 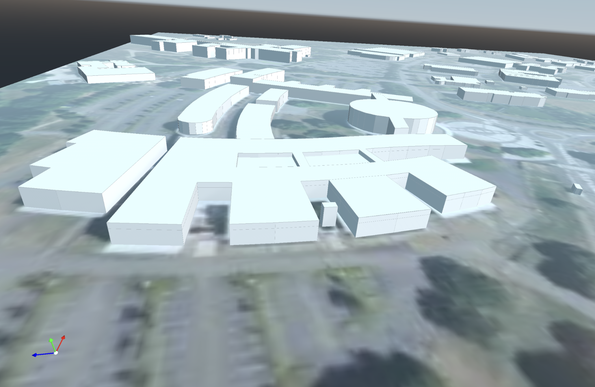}
  \label{fig:building-lod1}
}\\ 

\subfloat[LOD-2: a building in full detail using BIM. Note that LOD-2 and LOD-1 are mixed.]{%
  \includegraphics[width=0.48\linewidth]{\imagedir 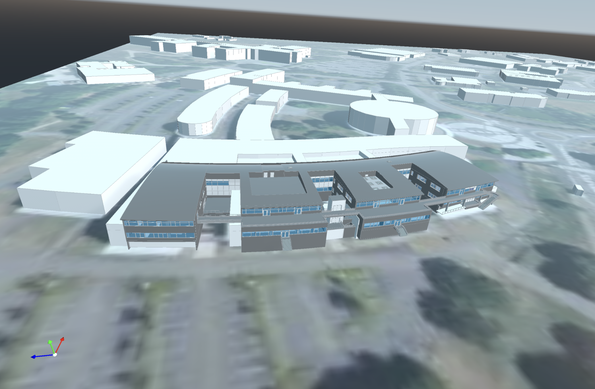}
  \label{fig:building-lod2}
}
\subfloat[LOD-2: A zoom on the LOD-2 building.]{%
  \includegraphics[width=0.48\linewidth]{\imagedir 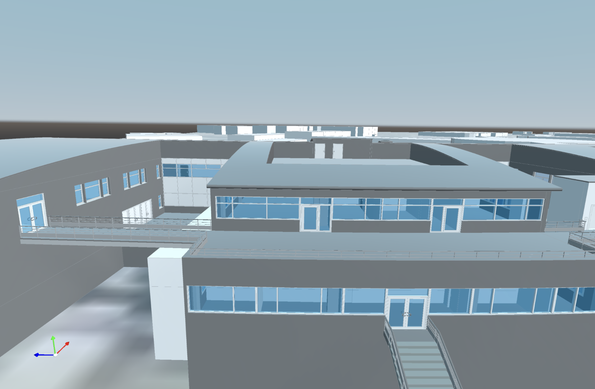}
  \label{fig:building-lod2-zoom}
}

\caption{Different representations of a building using our LOD definition}
\label{fig:buildings}
\end{wrapfigure}
\begin{wrapfigure}{R}{0.6\textwidth}
\centering
\subfloat[LOD-0\label{fig:city-strasbourg-lod0}]{%
  \includegraphics[width=0.4\textwidth]{\imagedir 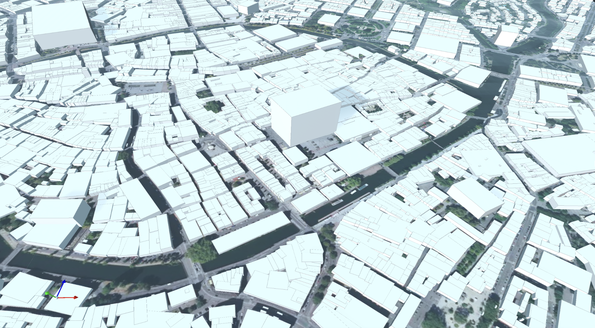}
}
\hfill 
\subfloat[LOD-1\label{fig:city-strasbourg-lod1}]{%
  \includegraphics[width=0.4\textwidth]{\imagedir 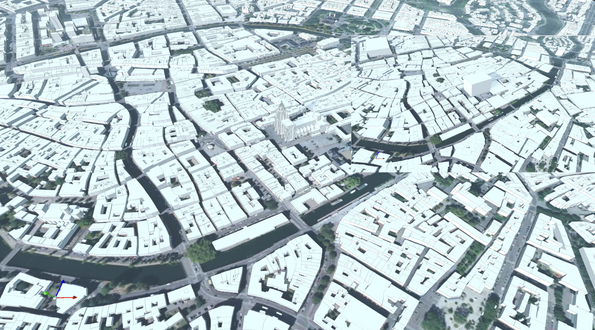}
}\\ 

\subfloat[LOD-1 terrain\label{fig:city-grenoble-terrain}]{%
  \includegraphics[width=0.4\textwidth]{\imagedir 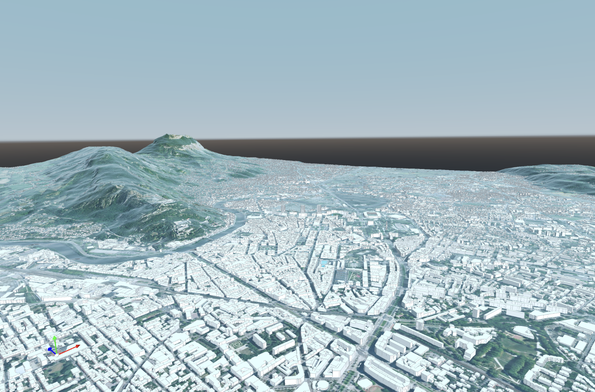}
}

\caption{Various representations of cities and terrain. Representation of Strasbourg center with LOD-0 in the top left panel and LOD-1 in the top right. LOD-1 city (Grenoble, France) representation with terrain elevation.}
\label{fig:city-strasbourg}
\end{wrapfigure}

In the figure~\ref{fig:buildings}, we illustrate the different levels of details. Panel~\ref{fig:building-lod0} displays the LOD-0 of a building with its bounding box.
Panel~\ref{fig:building-lod1} displays the LOD-1 of a building using its footprint elevated to its height. 
Panel~\ref{fig:building-lod2} and~\ref{fig:building-lod2-zoom} display the LOD-2 representation using BIM. In the figure~\ref{fig:city-strasbourg}, we display an illustration of the center of the city of Strasbourg with LOD-0, see panel~\ref{fig:city-strasbourg-lod0}, and LOD-1, see panel~\ref{fig:city-strasbourg-lod1}, representations.

\subsubsection{Building Modeling}
Building meshes are generated from metadata fetched from web services like OpenStreetMap~\cite{openstreetmap_contributors_planet_2017}, which provides multi-polygons with holes representing the complex urban fabric. For \textbf{LOD-0 and LOD-1}, buildings are modeled from 2D footprints extruded to form three-dimensional volumes. These volumes are then combined or subtracted to represent the district or the entire city, applying union operations on buildings that touch or intersect. In \textbf{LOD-2}, the focus shifts to creating conformal and watertight meshes suitable for detailed simulation tasks. These meshes are generated from Building Information Modeling (BIM) data in IFC format, enabling a detailed representation of each building component.

\subsubsection{Terrain Modeling}
The terrain modeling process utilizes elevation data extracted from raster images. The initial step involves creating a uniform mesh based on the size of the raster image. Following this, the elevation at each node is evaluated.
However, this method provides a very refined mesh, even when the terrain is flat. For this reason, we plan to apply the following procedure:
\begin{itemize}
    \item Compute an arbitrary number of isolines for elevation,
    \item Build a new terrain mesh conform with the isolines nodes and adapt to terrain elevation gradient.
\end{itemize}

\subsubsection{Vegetation Modeling}
Including vegetation (trees) in this geometric model is essential for the simulation of buildings. The shading provided by trees and the cooling provided by wooded areas has a significant impact. 
We use the OpenStreetMap database to obtain the metadata associated with the vegetation. In addition to the position of the trees, we can also obtain other information, such as the tree's height and species. The current strategy is as follows:
\begin{itemize}
\item Definition of a reference tree library that lists a certain number of tree species at different levels of precision (LOD). These geometric entities are parameterized so that a transformation can be applied.
\item Fetching vegetation metadata via OpenStreetMap. The attributes (height, species) are not always available, so we try to define them as best we can (by searching around).
\item Creating a tree geometric model using an affine transformation of the reference tree model.
\end{itemize}

\subsubsection{Integration of all urban geometric components}
This process involves creating a conforming mesh that includes both components (buildings, terrains, vegetation). Moreover, this step requires accomplishing some challenges by using complex geometric algorithms to realize the following points: \textit{(i)} Ensure that buildings on slopes are accurately modeled by adapting their height and embedding them into the terrain mesh. The figure~\ref{fig:city-grenoble-terrain} illustrates this aspect; \textit{(ii)} applies the intersection of buildings and consequently defines contact zones that can be included in building thermal models (coupling). \textit{(iii)} Apply the intersection of vegetation with buildings and terrain. 
And \textit{(iv)} Improve mesh quality after the previous geometric operations (mesh adaption).

\subsubsection{Visual Representation}
Advanced rendering techniques visualize the multi-fidelity urban models, supporting detailed analysis and general urban planning discussions. These visualizations are crucial for assessing the impact of urban changes and for stakeholder engagement.

This approach enhances urban models' accuracy, utility, and scalability, making them vital for comprehensive urban analysis and planning in the HiDALGO2 project.

\subsubsection{Computational Tools for Meshes Generation}

In the urban modeling process, particularly in generating building meshes, the Computational Geometry Algorithms Library (CGAL) plays a pivotal role. CGAL, \cite{the_cgal_project_cgal_2024} is renowned for its robust and efficient algorithms, which are crucial for handling complex geometric data and generating high-quality meshes. This library is currently used for the following operations: boolean operation of polygon, multi-polygon repairs, mesh generation, mesh intersection, mesh adaptation, and building roof skeleton.

All the features described above are not yet operational in our framework. We can reconstruct a geometric model of an urban area whose location is arbitrary so that the user can select an area anywhere in the world. The final mesh comprises all buildings LOD-0 or LOD-1 and the terrain with elevation. The integration of all the urban components is currently being investigated. This part requires fairly costly algorithms (mesh intersection), for which it will be necessary to form groups intelligently to reduce the computational cost.  

\paragraph{Current Mesh Generation Strategy}
The current strategy for mesh generation employs multi-threading (MT) to handle various stages of the mesh construction process. Initially, the tiles used are determined based on the specified location, a task performed sequentially. Following this, GIS data, including buildings and elevation information, is downloaded in parallel. Next, polygons are repaired to ensure they are suitable for mesh generation, with this process executed in parallel using MT. The terrain mesh is then generated in parallel, employing CGAL's algorithms to ensure precision and efficiency. Union operations at tile junctions are performed to ensure continuity, a step that is currently sequential. Finally, building meshes are generated using a parallel MT approach, leveraging CGAL for its advanced mesh generation capabilities.

\paragraph{Advancing Towards Full Parallelism}
The following steps in enhancing the mesh generation process involve moving towards a fully parallel approach using both Multi-threading (MT) and Message Passing Interface (MPI). The goal is to scale the process to handle entire cities or larger urban areas. This scale-up involves utilizing MPI to manage distributed computing resources effectively. To achieve this, partitions of tiles with overlapping regions are created to ensure complete and accurate building descriptions without requiring extensive MPI communications. Furthermore, each process uses an MT strategy to generate meshes independently, with overlapping zones allowing for the creation of complete building structures.

\subsubsection{Partitioning Strategies Depending on Simulation Use Cases}
Data partitioning is a crucial stage in the deployment of the supercomputer simulator. We need to partition the geometry of the city in such a way as to distribute the simulation computation correctly.
Different partitioning strategies are considered depending on the specific requirements of the simulation use cases :
\begin{itemize}
\item Case 0: simple scenarios where buildings do not interact with their environment, a basic listing and weighting strategy is sufficient
\item Case 1: buildings interact with environmental elements, a more complex partitioning strategy is necessary that considers both the buildings and their immediate non-building surroundings. The build meshes and environment meshes (terrain, vegetation) are integrated conformably; each component can be partitioned separately.
\item  Case 2: for full interaction models, the partitioning strategy starts with the buildings and extends to the entire urban mesh, ensuring all elements are considered to minimize communication overhead.
\item Case 3: Regarding scenarios with extreme partitioning needs, Test Case 3 employs a multi-grid approach, defining coarse and fine meshes to manage computational resources efficiently. This multi-fidelity approach enhances the accuracy and applicability of urban models in simulations and ensures scalability across different computational platforms, making it a cornerstone of urban analysis.
\end{itemize}

The current version of our simulation framework has implemented case 0 and case 1. Cases 2 and 3 require a full conformal mesh, which is not yet available. Moreover, these partitioning strategies are costly, so we have planned to use other third-party tools, such as Zoltan2~\cite{the_zoltan2_team_zoltan2_nodate}, to improve efficiency.
\begin{wrapfigure}{R}{0.6\textwidth}
\centering
\subfloat[Partitioning Case 0\label{fig:city-strasbourg-lod0-parts}]{%
  \includegraphics[width=0.48\linewidth]{\imagedir 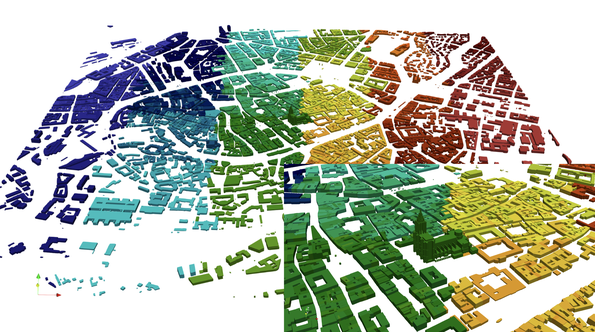}
}
\subfloat[Partitioning Case 1\label{fig:city-strasbourg-lod1-parts}]{%
  \includegraphics[width=0.48\linewidth]{\imagedir 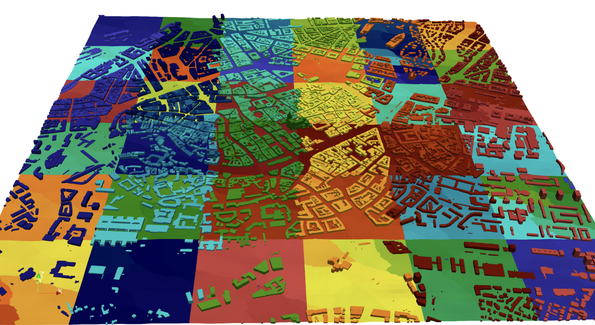}
}
\caption{Mesh partitioning illustrations}
\label{fig:partitioning}
\end{wrapfigure}

\begin{figure}[htbp]
\centering
\subfloat[View on whole 3D mesh\label{fig:city-ny-largescale-whole}]{%
  \includegraphics[width=0.45\textwidth]{\imagedir 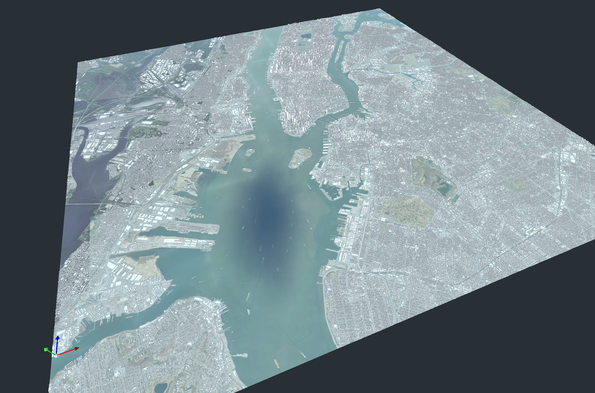}
}
\hfill
\subfloat[Zoom on Manhattan buildings\label{fig:city-ny-largescale-zoomB}]{%
  \includegraphics[width=0.45\textwidth]{\imagedir 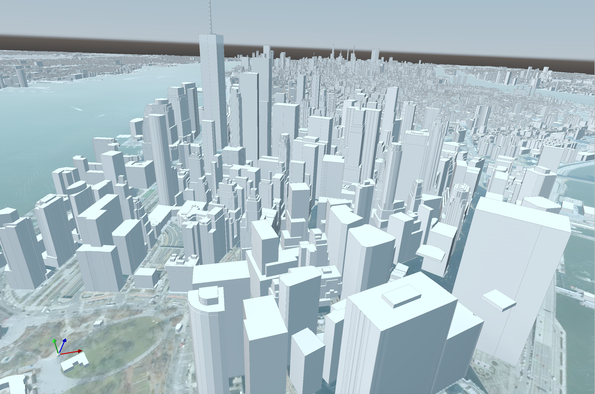}
}\\
\subfloat[Focus on Manhattan\label{fig:city-ny-largescale-zoomZ}]{%
  \includegraphics[width=0.45\textwidth]{\imagedir 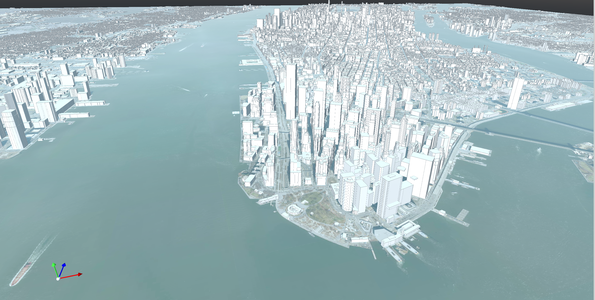}
}
\hfill
\subfloat[Central Park\label{fig:city-ny-largescale-zoomC}]{%
  \includegraphics[width=0.45\textwidth]{\imagedir 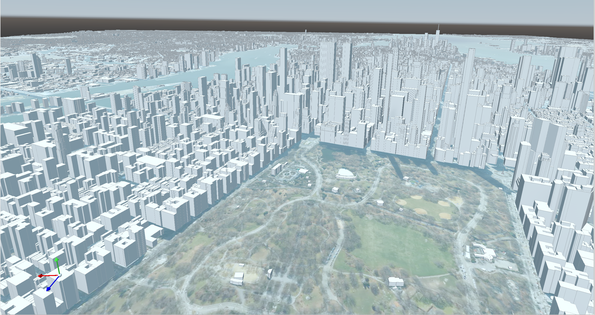}
}
\caption{$20 \times 20 \mathrm{km}^2\ $ geometric reconstruction of New York City (LOD-1)}
\label{fig:city-ny-largescale}
\end{figure}

The figure~\ref{fig:partitioning} illustrates the different strategies discussed previously and presents a reconstruction of New York City in figure \ref{fig:city-ny-largescale}. This geometric model has an area of 400 $km^2$ (20 km square side of 20 km) and has generated around 450000 buildings. This example requires the large-scale approach discussed but not yet implemented or tested. 

\subsubsection{Conclusion on mesh construction}
Our strategy aims to improve the efficiency and scalability of the mesh generation process and enhance the fidelity and accuracy of the urban models used in simulations.

\subsection{Modeling and simulations}

We now turn to describing the physical modeling of city energy simulation.
To do this, we built on two main tools: Feel++ and Modelica. 

\subsubsection{KUB is based on the Feel++ toolchain}

Feel++ is a comprehensive framework designed to tackle problems based on Ordinary Differential Equations (ODEs) and Partial Differential Equations (PDEs). Using modern C++ (C++17 and C++20) standards coupled with a Python layer through Pybind11, Feel++ enables seamless parallelism and is equipped with default communicators that simplify handling complex computational tasks. The framework's versatility is evident in its deployment across various platforms, including research and educational environments and cloud services tailored for high-performance computing needs.

Key features of the Feel++ framework encompass an extensive range of numerical methods designed to address Partial Differential Equations (PDEs), see~\cite{christophe_prudhomme_feelppfeelpp_2024}. These methods include continuous Galerkin (cG), discontinuous Galerkin (dG), hybrid discontinuous Galerkin (hdG), and reduced basis methods (rb/mor). A Domain Specific Language (DSL) for Galerkin methods significantly enhances the ease of implementing and experimenting with new numerical methods. The de Rham complex provides a comprehensive toolkit for constructing finite element spaces of arbitrary order, facilitating precise mathematical modeling. The framework's automatic differentiation and symbolic integration capabilities effectively bridge the gap between mathematical expressions and their computational implementation. Feel++ supports diverse applications, from fluid dynamics and structural mechanics to heat transfer and electrostatics, demonstrating its flexibility and broad applicability. Furthermore, its integration with Specx for task-based parallel execution optimizes performance and scalability on modern computational architectures.


Documentation and further details can be accessed through Feel++ Toolboxes Documentation\footnote{\url{https://docs.feelpp.org/toolboxes/latest/}}.
This powerful toolchain is essential for KUB.

\subsubsection{Computing Shading Masks and View Factors with Feel++}

In city energy simulations, the computation of shading masks and view factors is crucial for accurately modeling the impact of solar radiation on building surfaces. Shading masks quantify the percentage of blocked solar radiation for each building surface (including walls and roofs) depending on the sun's direction. This is influenced by nearby structures such as other buildings, vegetation, and urban furniture. The view factors describe the fraction of radiation that leaves one surface and strikes another, essential for calculating radiative heat exchanges between building surfaces.

\paragraph{Numerical Methods and Challenges}
Both shading masks and view factors are computed using Monte Carlo and ray tracing techniques, which allow for handling complex geometries with various obstructions. Despite being purely geometric quantities, these calculations face significant challenges such as:
\begin{itemize}
    \item Efficient computation of integrals for view factors, especially when considering specular surfaces that require multiple ray bounces.
    \item Managing large-scale mesh computations and data storage, particularly when detailed urban environments are modeled.
\end{itemize}

\paragraph{Implementation in Feel++}
Feel++ facilitates these computations through its robust numerical methods optimized for high performance and parallel execution. For each face of a building, Feel++ computes solar masks using a Monte Carlo approach for various sun positions, ensuring efficient and scalable processing across multiple CPU cores. This enables the integration of dynamic solar shading effects into the simulation of building energy performance, providing a more accurate representation of real-world conditions.

\begin{figure}[htbp]
\centering
\subfloat[LOD-0\label{fig:sm-building-east}]{%
  \includegraphics[width=0.45\textwidth]{\imagedir 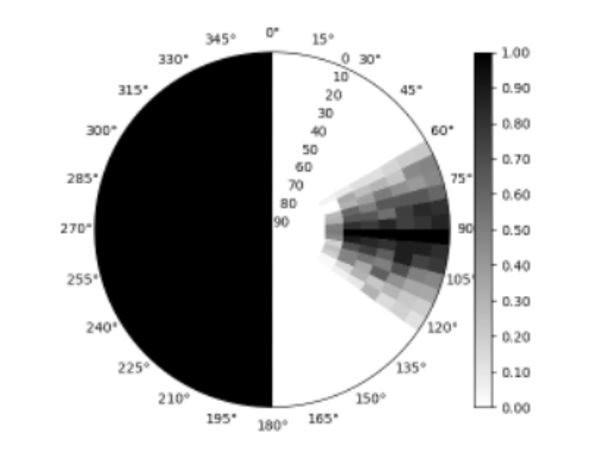}
}
\hfill
\subfloat[LOD-1\label{fig:sm-whole-building}]{%
  \includegraphics[width=0.45\textwidth]{\imagedir 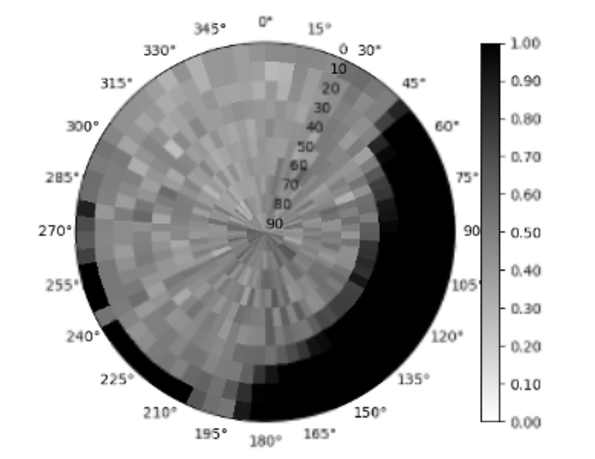}
}\\
\subfloat[LOD-1 Large scale\label{fig:sm-strasbourg}]{%
  \includegraphics[width=0.45\textwidth]{\imagedir 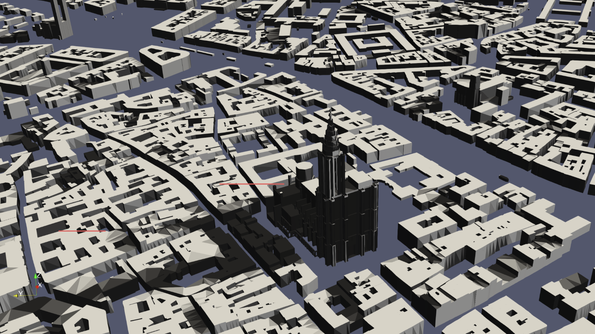}
}
\hfill
\subfloat[Heat transfer benchmark in 2D including view factors~\cite{van_eck_surface_2016}\label{fig:view-factor}]{%
  \includegraphics[width=0.45\textwidth]{\imagedir 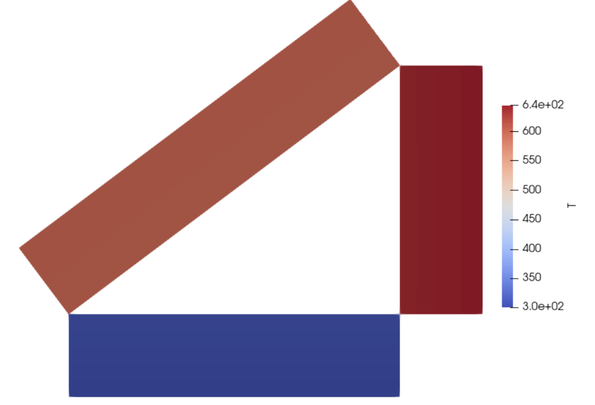}
}
\caption{Solar masks and view factors computations}
\label{fig:solar-masks-vf}
\end{figure}

In the figure~\ref{fig:solar-masks-vf}, we illustrate \textit{(i)} the solar masks of building face oriented eastwise for a discretization of the sun position in the panel~\ref{fig:sm-building-east}, \textit{(ii)} the solar masks for an entire building for discretization of the sun position in the panel~\ref{fig:sm-whole-building}, \textit{(iii)} a visualization of the solar mask early morning of the city center of Strasbourg in panel~\ref{fig:sm-strasbourg} and \textit{(iv)} an image of a standard benchmark~\cite{van_eck_surface_2016} for the computation of view-factors and solving the heat transfer problems between three building blocks in 2D subsequently in panel~\ref{fig:view-factor}.

\subsubsection{Heat Transfer Modeling with Feel++ and Modelica}
In the Urban Building Energy Model (UBEM), heat transfer analysis is crucial for predicting energy consumption, internal air temperature variations, and overall building energy performance. This analysis is facilitated by a combination of Feel++ and Modelica, enabling detailed simulations of heat dynamics within urban buildings.

\paragraph{Modelica for Multizone Heat Transfer}
Modelica offers extensive capabilities for multizone building energy simulations, utilizing models that range from simple (LOD-0) to more complex (LOD-1) representations. The multizone approach in Modelica is beneficial for modular and scalable simulations, where each building zone can be modeled with different fidelity based on the simulation requirements. This method leverages the generation of Functional Mock-up Units (FMUs), which integrate seamlessly into larger C/C++ applications, providing a robust framework for handling complex simulations involving multiple interacting systems.

\paragraph{Finite Element Analysis with Feel++}
Complementing Modelica's capabilities, Feel++ provides robust tools for finite element analysis, particularly in handling the detailed aspects of heat transfer within urban environments. It uses advanced numerical methods like reduced basis methods for rapid scenario testing and parallel-in-time algorithms for efficient simulations. This is particularly important for assessing the impact of solar radiation and external shading, which are modeled using geometric and dynamic shading masks derived from solar paths.

\paragraph{Integrated Approach}
The integration of Feel++ and Modelica is exemplified in their use of shading masks and view factors, which are critical for accurate solar heat gain calculations. These masks are computed using Monte Carlo simulations and ray tracing methods to assess the percentage of solar radiation impacting various building surfaces. This data feeds into the Modelica simulations, enhancing the accuracy of the thermal load predictions.

\paragraph{Challenges and Solutions}
One of the main challenges in urban building simulation is managing the computational load, which is addressed through computing strategies that leverage currently CPUs but in the GPUs but, in the future, will enable hybrid computing with both CPUs and GPUs. This approach ensures that large-scale simulations, necessary for city-wide energy analysis, remain feasible and efficient. Additionally, the mesh partitioning techniques discussed earlier are employed to optimize the data handling and processing times, further integrating the spatial data management with the thermal modeling processes.





\section{CI/CD Framework for the Urban Building Pilot}
The Urban Building pilot utilizes the Feel++ framework, supported by a robust CI/CD framework that facilitates efficient development and deployment. 

\subsection{Standard CI/CD DevOps}
The development and deployment of KUB builds on top of the Feel++ CI/CD framework. It employs GitHub Actions and Docker: GitHub Actions automate real-time workflows to compile, test, and validate code changes, facilitating rapid development cycles and ensuring code quality. On the other hand, Docker provides a containerized environment that encapsulates Feel++ and its dependencies, ensuring consistent operations across diverse computing environments. These Docker images, customized for various system requirements, are maintained on the GitHub Container Registry (ghcr.io) to accommodate multiple deployment scenarios.

\begin{figure}
    \centering
    \includegraphics[width=\textwidth]{\imagedir 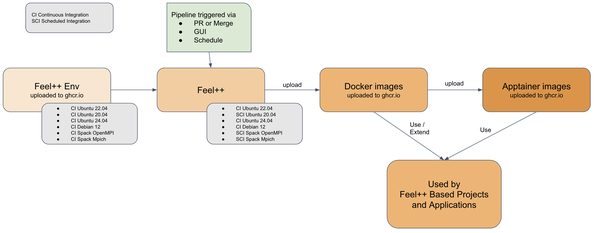}
    \caption{CI/CD DevOps for Feel++}
    \label{fig:feelpp-devops}
\end{figure}
The CI/CD workflow, see Figure~\ref{fig:feelpp-devops}, is crucial for efficiently integrating and deploying updates across all projects that utilize the Feel++ framework. The workflow leverages various main ingredients of GitHub Actions features.
\begin{enumerate}
    \item \textbf{Pull Requests and Merges:} Triggering CI to verify that new code integrations meet all tests and standards.
    \item \textbf{Graphical User Interface (GUI) thanks to \texttt{workflow\_dispatch}:} Enabling developers to manually trigger pipelines through a GUI, which facilitates rapid deployment or testing. 
    \item \textbf{Scheduled Runs:} Conducting regular updates and maintenance checks to ensure continuous system integrity and responsiveness.
\end{enumerate}

\subsection{HPC DevOps (HPCOps)}
Feel++ CI/CD workflow for high-performance computing applications incorporates specialized HPCOps (HPC DevOps) practices that ensure the software performs consistently across various HPC systems. 

\begin{figure}
    \centering
    \includegraphics[width=\textwidth]{\imagedir 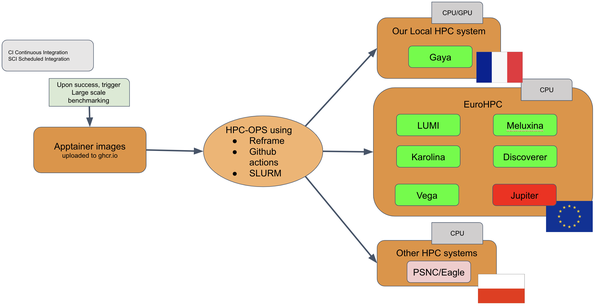}
    \caption{CI/CD HPCOps for Feel++}
    \label{fig:feelpp-hpcops}
\end{figure}

Figure~\ref{fig:feelpp-hpcops} illustrates the HPC CI/CD or HPCOps workflow for Feel++. 

The tools and strategies for HPCOps are  \textit{(i) }\textbf{Reframe-HPC:} Utilized to define and manage systematic benchmarks that are reproducible across different HPC environments, facilitating the testing of performance and scalability, see~\cite{karakasis_reframe-hpcreframe_2024};
\textit{(ii)}\textbf{SLURM:} Employs its REST API if available, \textit{e.g.} on MeLuXiNa, otherwise scripted SLURM usage for CI/CD for scheduling and managing jobs on integrated HPC systems, allowing programmable job submission and monitoring directly from CI workflows, see~\cite{slurm_development_team_slurm_2024}; and \textit{(iii)} \textbf{Apptainer:} Ensures that Docker containers can be deployed securely and efficiently in HPC settings, supporting portability and consistency, see~\cite{apptainer_contributors_apptainer_2024}.

Integration with EuroHPC JU supercomputers such as LUMI, Karolina, Meluxina, Discoverer, Vega, and Leonardo enhances the capability to perform large-scale simulations and check the parallel properties and correctness of Feel++. The operations include automatic testing that triggers larger-scale tests on designated HPC nodes once new changes are integrated and verified by standard CI/CD pipelines.

Regarding monitoring and reporting, performance results from these operations are automatically captured and uploaded to the data storage system, such as the performance reports 

\begin{wrapfigure}{R}{.7\linewidth}
    \centering
    \includegraphics[width=\linewidth]{\imagedir 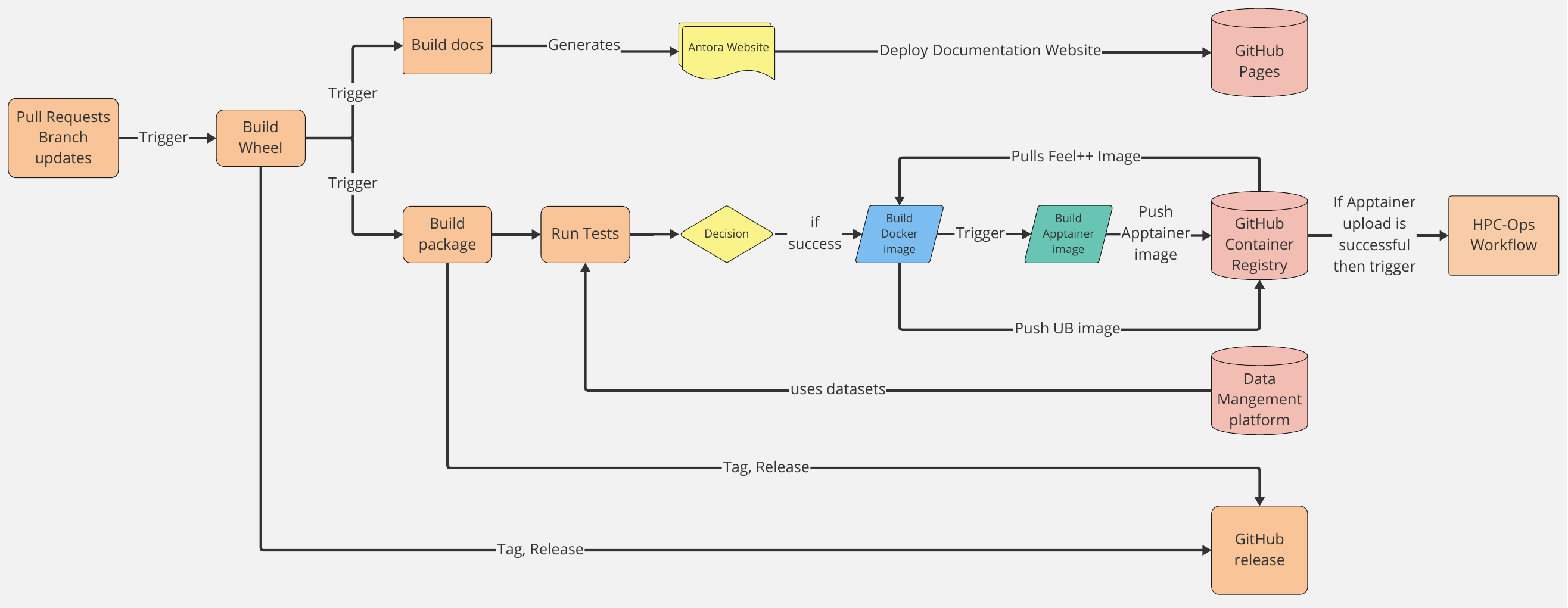}
    \caption{KUB standard DevOps}
    \label{fig:kub-devops}
\end{wrapfigure}
This framework leverages cutting-edge computational technologies, ensuring the high performance and accuracy of the Feel++ framework and Feel++-based applications. 
It sets a benchmark for integrating modern software frameworks with advanced HPC infrastructure to significantly advance computational research and applications.

Figure~\ref{fig:kub-devops} shows that the CI/CD standard DevOps framework for KUB uses similar steps.

\begin{figure}
    \centering
    \includegraphics[width=\textwidth]{\imagedir 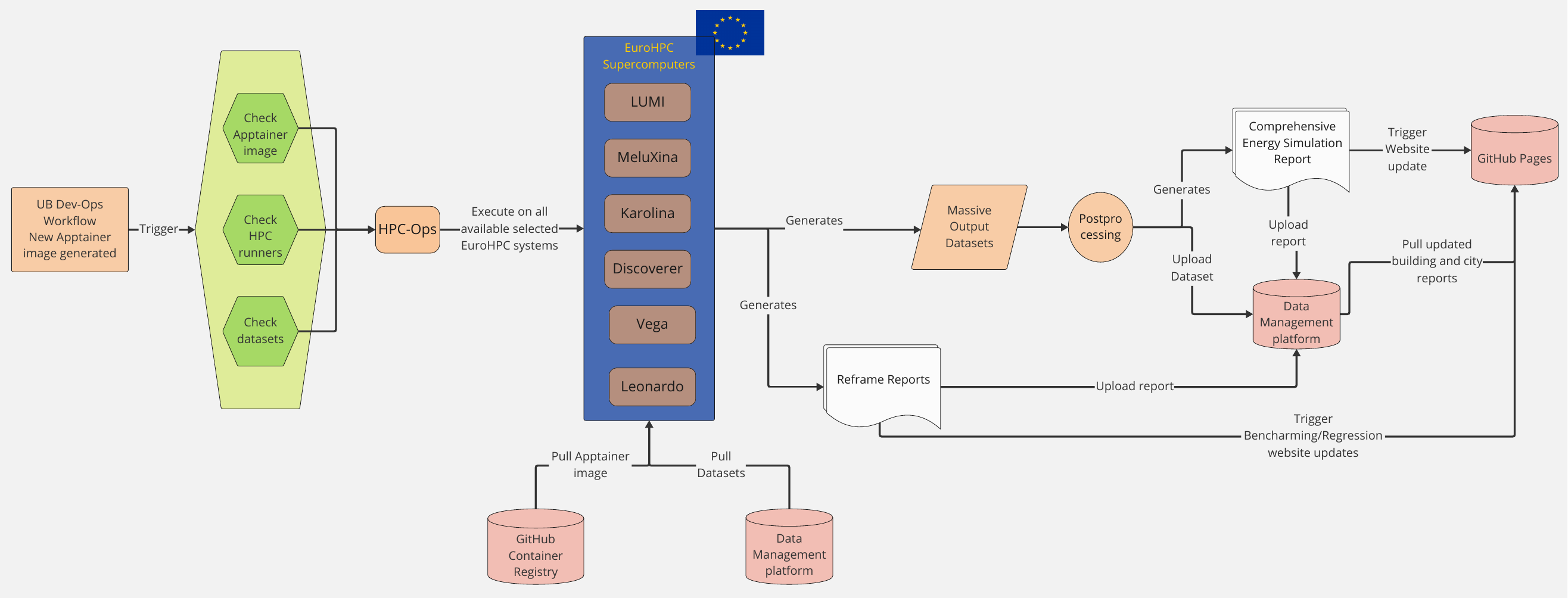}
    \caption{Ktirio Urban Building HPCOps workflow}
    \label{fig:kub-hpcops}
\end{figure}

Figure~\ref{fig:kub-hpcops} finally shows the HPCOps deployed on EuroHPC systems to check the parallel properties and correctness of the pilot on large-scale cases. It extends Feel++ HPC ops with additional steps to handle our complex pre- and post-processing steps.

\subsection{Benchmarking KUB}

\begin{wrapfigure}{R}{0.6\textwidth}  
  \centering
  \begin{subfloat}[Scalability tests on a few EuroHPC systems from 1 to 32 nodes of 128 cores per node]{
    \includegraphics[width=0.9\linewidth]{\imagedir 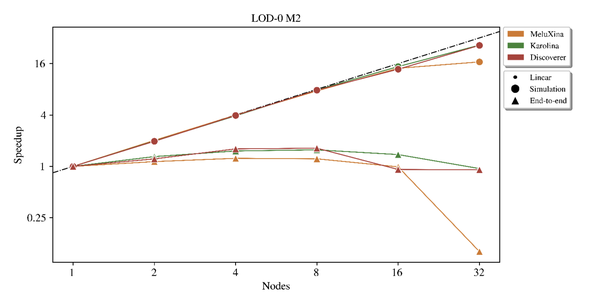}
    \label{fig:scalability}
  }\end{subfloat}
  \\
  \begin{subfloat}[Execution breakdown on a few EuroHPC systems from 1 to 32 nodes of 128 cores per node]{
    \includegraphics[width=0.9\linewidth]{\imagedir 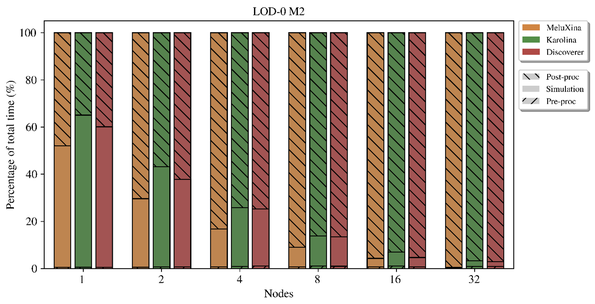}
    \label{fig:execution-breakdown}
  }\end{subfloat}
  \caption{Detailed performance metrics for scalability and execution on EuroHPC systems}
  \label{fig:combined-metrics}
\end{wrapfigure}

Finally, we display some results of our benchmarking activities~\cite{hidalgo2_d31_2024} for the KUB application regarding HPC performance. Running these experiments regularly is essential for maintaining the program's efficiency when developing the application and after machine updates. 
Hence, we have used the KUB application in the Strasbourg city center, with a square area of $4 \mathrm{km}^2$ (approximately 17K buildings). The scalability results were realized on EuroHPC JU systems obtained with our HPCOps pipeline; see Figure~\ref{fig:kub-hpcops}. Figure \ref{fig:scalability} depicts the speedup achieved on Discoverer, Karolina, and MeluXina. They present results for the total execution (end-to-end) of the KUB application and the simulation component.

The pipeline's simulation part scales almost linearly, which is expected as the buildings are not coupled together in the model currently used for the simulation. On the other hand, the total execution of the pilot's pipeline does not scale; as more nodes are employed, the performance degrades.

To better understand what causes this degradation, we measure the computing times of the different stages of the pipeline and present the execution breakdown in Figure \ref{fig:execution-breakdown}. This figure reports the portion of the total execution taken by:
\begin{itemize}
\item Pre-processing (Pre-proc): The time elapsed in initialization before entering the time loop of the simulation
\item Simulation (Simulation): The cumulative time spent calculating the new solution at each time step
\item Post-processing (Post-proc): The cumulative time spent exporting results, i.e., generating files containing the output of the UB model.
\end{itemize}

Pre-processing does not scale. However, it occupies only a small part of the total execution, and thus it is not performance-critical. On the other hand, as more nodes are employed and the time spent in the actual simulation is decreased, the post-processing stage dominates the execution. It becomes the main bottleneck, causing the previously observed performance degradation.
This behavior is caused by the multiple files being written in parallel on the shared file system. More specifically, most of the writing time is spent in opening and closing files in parallel. We are investigating potential solutions, such as asynchronous writes, data caching, etc. Finally, as the project progresses, we expect the urban building models used in the simulation to become more complex, leading to an increase in the time occupied by the simulation part and, hence, to a reduction of the impact of post-processing on the total execution time of the pilot.

\section{Conclusion}

We are developing the computational Ktirio Urban Building(KUB) framework: assembling this very compelling application encompasses challenges in mathematics --- scalable modeling and simulation, large-scale watertight robust mesh generation, advanced analysis including data simulation --- and computer science --- scalable framework, software architecture, modern development, testing and packaging environment including standard DevOps and now HPCOps.---
The overall programming, integration, delivery, and deployment environment is critical to develop such an application. To our knowledge, this is the first application that can be automatically benchmarked and executed on EuroHPC JU systems thanks to CI/CD or HPCOps. Moreover, the workflow from Feel++ to KUB provides a considerable gain in terms of development and testing time, automated as much as possible, and enabling researchers and developers to focus their work better.

Our next steps include \textit{(i)} enabling tasks-based parallelism using the C++ framework Specx, see~\cite{cardosi_specx_2022}; \textit{(ii)} improving the modeling and simulation components, including mesh generation, handling of vegetation and urban furniture, and enabling view factors as well as providing a variety of configurable building energy models; \textit{(iii)} enhanced parallel strategies particularly in terms of partitioning and improved large scale I/O; and of course \textit{(iv)} pursue our benchmarking activities on EuroHPC JU supercomputers.

\begin{credits}
\subsubsection{\ackname} Funded by the  European  Union. This work has received funding from the European High-Performance Computing Joint Undertaking (EuroHPC JU)  and  Poland, Germany,  Spain,  Hungary,  France, and Greece under grant agreement number 101093457. This  publication  expresses  the  opinions  of  the  authors  and  not  necessarily those of the EuroHPC JU and Associated Countries which are not responsible for any use of the information contained in this publication

Part of this work was also funded by \textit{(i)} the France 2030 NumPEx Exa-MA (ANR-22-EXNU-0002) project managed by the French National Research Agency (ANR), \textit{(ii)} AMIES, the french agency for interaction between mathematics and enterprises and \textit{(iii)} CNRS through its prematuration programme.

We acknowledge the EuroHPC Joint Undertaking for awarding this project access through  EuroHPC Development Access grants EHPC-DEV-2024D05-025 and EHPC-DEV-2023D08-047 to the EuroHPC JU supercomputers : \textit{(i)} Kumi, hosted by CSC (Finland) and the Lumi consortium, \textit{(ii)} MeluXina hosted by LuxProvide, Luxembourg, \textit{(iii)} Karolina hosted by IT4Innovations National Supercomputing Center, Czechia, \textit{(iv)} Discoverer hosted by Sofia Tech Park, Bulgaria, \textit{(v)} Vega hosted by IZUM, Slovenia, and \textit{(vi)} Leonardo hosted by CINECA, Italy.

Finally the authors would like to acknowledge the many fruitful discussions with our partners Luc Kern from Synapse Concept and Leopold Fischer from Cisco Meraki, our colleagues \textit{(i)} from Hidalgo2 ICCS Kostis Nikas, Aristomenis Theodoridis and Petros Anastasiadis for the discussions on Reframe and joining their EuroHPC access grant, \textit{(ii)} Hidalgo2 HLRS Sameer Haroon for the discussions on CI/CD, \textit{(iii)} Pierre Alliez from INRIA Titane and Andreas Fabri from Geometry Factory regarding the discussions on CGAL and using Polygon Repair, and finally \textit{(iv)} our former colleague Zohra Djatouti, now at Kipsum, which whom we initiated this endeavor. 

\end{credits}

\bibliographystyle{splncs04}
\bibliography{references}

\end{document}